# QUADRATURE COMPRESSIVE SAMPLING SAR IMAGING

*Huizhang Yang, Shengyao Chen, Feng Xi and Zhong Liu*

Department of Electronic Engineering
Nanjing University of Science & Technology
Nanjing, Jiangsu, 210094, P. R. China

**ABSTRACT**

This paper presents a quadrature compressive sampling (QuadCS) and associated fast imaging scheme for synthetic aperture radar (SAR). Different from other analog-to-information conversions (AIC), QuadCS AICs using independent spreading signals sample the SAR echoes due to different transmitted pulses. Then the resulting sensing matrix has lower correlation between any two columns than that by a fixed spreading signal, and better SAR image can be reconstructed. With proper setting of the spreading signals in QuadCS, the sensing matrix has the structures suitable for fast computation of matrix-vector multiplication operations, which leads to the fast image reconstruction. The performance of the proposed scheme is assessed using real SAR image. The reconstructed SAR images with only one-fourth of the Nyquist data achieve the image quality similar to that of the classical SAR images with Nyquist samples.

## 1. INTRODUCTION

Synthetic aperture radar (SAR) with compressed sensing (CS) has received wide attention in recent years. Different from classical SAR imaging, CS SAR operates on sub-Nyquist data and reconstructs its images under the assumption of sparse targets in a certain domain [1]. CS SAR acquires SAR data at sub-Nyquist rate and thus requires low data storage and low bandwidth data link. In addition, SAR also benefits from CS in terms of sidelobe suppression and wider swaths.

In a practical CS SAR system, the sub-Nyquist data in range dimension is often acquired by analog-to-information conversion (AIC) and the data reduction in azimuth dimension is realized through transmitting a small number of pulses during a coherent processing interval (CPI). Several AICs have been suggested for sampling radar echoes at range dimension, such as random sampling [2], RMPI [3], Xampling [4] and QuadCS [5]. In this paper, we consider SAR data acquisition by QuadCS AIC and its image reconstruction.

This work was supported in part by National Science Foundation of China under Grants 61571228, 61671245.

In the reported AIC SARs, a fixed AIC is taken to sample the radar echoes [6, 7] as an analog-to-digital converter (ADC). The 'fixed' here means that the AIC sampling the radar echoes due to different transmitted pulses does not change. We adopt the QuadCS structure [5, 8] and develop the sampling of SAR echoes with inter-pulse independent spreading signals in QuadCS. Then the QuadCS AIC sampling the radar echoes is different from one pulse to another pulse. The sampling strategy will generate better sensing matrix than that by using a fixed spreading signal. It is well-known that the CS reconstruction performance greatly depends on the CS sensing matrix and the better sensing matrix leads to the better reconstruction performance [9]. Then the SAR using the proposed sampling strategy will have better SAR image reconstruction quality.

Moreover, with proper setting of spreading signals, the QuadCS sensing matrix has special structures suitable for fast computation of matrix-vector multiplication operations, which are common operations in most image reconstruction algorithms. Then a fast image reconstruction can be obtained. Simulations with SAR data generated from real SAR image demonstrate that the proposed scheme can reconstruct high-quality image and is superior to other related schemes.

## 2. SAR MODEL

SAR imaging model is well developed depending on the imaging algorithms [9]. In this paper, we take the SAR imaging model based on the chirp-scaling algorithm (CSA) as an example for our discussion.

Assume that the SAR works in stripmap mode and travels along a well-defined path. The transmitting waveform is defined by $\mathrm{Re}\{s(t)e^{j2\pi f_0 t}\}$, where $s(t)$ is the complex baseband pulse with the bandwidth $B$, $f_0$ is the carrier frequency and $\mathrm{Re}\{\bullet\}$ denotes the real part of $\{\bullet\}$. Let $\mathbf{v}$ be the spatial vector of illuminated scene and $\mathbf{u}_l$ be the radar position at the $l$-th observation. Then the received echo backscattered from the illuminated scene with reflection function $z(\mathbf{v})$ is given by

$$y_l(t) = \mathrm{Re}\left\{ e^{j2\pi f_0 t} \int z(\mathbf{v}) \exp\left(-j4\pi f_0 \|\mathbf{v}-\mathbf{u}_l\|_2 / c\right) \times s\left(t - 2\|\mathbf{v}-\mathbf{u}_l\|_2 / c\right) d\mathbf{v} \right\}. \quad (1)$$

Upon receiving, the echo (1) is down-converted to an intermediate-frequency (IF) signal or a baseband signal, by which IF sampling or baseband sampling is applied to obtain the complex baseband samples. In a pulse repetition interval $T$, we obtain $N_r$ Nyquist samples. Denote $\mathbf{y}_l$ as the length-$N_r$ sample vector. Then, in a coherent processing interval $N_a T$, we can formulate a two-dimensional SAR data as $\mathbf{Y} = [\mathbf{y}_1, \cdots, \mathbf{y}_{N_a}] \in \mathbb{C}^{N_r \times N_a}$. As discussed in [10], for the linear frequency modulated (LFM) transmitting waveform, the CSA-based SAR generates the SAR image $\mathbf{Z} \in \mathbb{C}^{N_r \times N_a}$,

$$\mathbf{Z} = \left( \mathbf{H}_3 \odot \mathbf{F}_r^H \left( \mathbf{H}_2 \odot \mathbf{F}_r \left( \mathbf{H}_1 \odot \mathbf{Y} \mathbf{F}_a \right) \right) \right) \mathbf{F}_a^H , \quad (2)$$

where matrices $\mathbf{F}_r$ and $\mathbf{F}_a \in \mathbb{C}^{N_r \times N_a}$ are normalized discrete Fourier transform (DFT) matrices, matrices $\mathbf{H}_1$, $\mathbf{H}_2$ and $\mathbf{H}_3 \in \mathbb{C}^{N_r \times N_a}$ are jointly used to perform range cell migration correction, range and azimuth compression and residual phase correction. The explicit expressions of these matrices are omitted here. In (2), $\odot$ and $(\cdot)^H$ denote the Hadamard product and the conjugate transpose operations, respectively. By reversing the process of the CSA as in [6], we can express $\mathbf{Y}$ as

$$\mathbf{Y} = \left( \mathbf{H}_1^* \odot \mathbf{F}_r^H \left( \mathbf{H}_2^* \odot \mathbf{F}_r \left( \mathbf{H}_3^* \odot \mathbf{Z} \mathbf{F}_a \right) \right) \right) \mathbf{F}_a^H , \quad (3)$$

where $(\cdot)^*$ denotes the conjugate operation. Let $\mathbf{y} = \text{vec}(\mathbf{Y}) \in \mathbb{C}^N$ and $\mathbf{z} = \text{vec}(\mathbf{Z}) \in \mathbb{C}^N$ be the column-wise vectors of matrices $\mathbf{Y}$ and $\mathbf{Z}$, respectively, where $N = N_r N_a$. Then (3) can be re-expressed as

$$\mathbf{y} = \mathbf{D} \mathbf{z} , \quad (4)$$

where $\mathbf{D}$ is an $N \times N$ matrix formulated from matrices $\mathbf{F}_r$, $\mathbf{F}_a$, $\mathbf{H}_1$, $\mathbf{H}_2$ and $\mathbf{H}_3$.

In CS SAR, we often obtain the sub-Nyquist data through randomizing the data $\mathbf{y}$. As shown in the next section, the QuadCS is one of randomizing techniques.

## 3. QUADCS SAR

In this section, we present our QuadCS SAR system. The sampling strategy is first introduced and then a fast imaging algorithm is discussed.

### 3.1. QuadCS SAR Sampling

The fundamentals of QuadCS AIC are shown in Fig. 1, which consists of a sub-Nyquist sampling subsystem and a digital quadrature demodulation subsystem. The first subsystem generates the low-rate samples of the input signal through random spreading, bandpass filtering and low-rate IF ADC. The second subsystem is used to extract digital compressive in-phase and quadrature components from the low-rate samples. Its operation is the same as in classic quadrature sampling. For notational convenience, we assume that the IF frequency is set to be $f_0$.

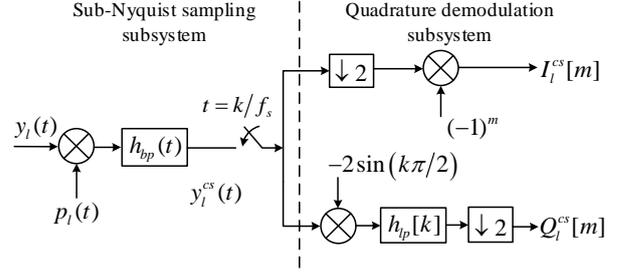

Fig. 1. The block diagram of QuadCS.

The random spreading is implemented through mixer operation which spreads the frequency content of the baseband signal to full spectrum of the spreading signal $p_l(t)$ with the bandwidth $B_p$ $(B_p \geq B)$. The bandpass filter $h_{bp}(t)$ is designed with the IF frequency $f_0$ and bandwidth $B_{cs}$. Because the bandwidth $B_{cs}$ can be set much smaller than the bandwidth $B$ $(B_{cs} \ll B)$, a low-rate IF ADC with the rate $f_s$ can be used to sample the compressive IF signal $y_l^{cs}(t)$. In a pulse repetition interval $T$, for Nyquist-rate IF ADC, the QuadCS generates a length-$M_r$ compressive sampling or measurement vector $\mathbf{y}_l^{cs} \in \mathbb{C}^{M_r}$, where $M_r = \lfloor B_{cs} T \rfloor$ and $\lfloor \cdot \rfloor$ is a floor function.

When applied to SAR sampling, the QuadCS AIC assumes inter-pulse independent random spreading signals. In our discussion, the spreading signal is assume to be of $T$-period. Denote the spreading signal as $p_l(t)$ for the $l$-th observation. We now derive the frequency-domain measurement model for the QuadCS SAR system.

Let us express the spreading signal $p_l(t)$ in its Fourier series,

$$p_l(t) = \sum_{i=-L_p}^{L_p} \rho_{i,l} e^{j 2\pi f_p i t} , \quad (5)$$

where $f_p = 1/T$, $\rho_{i,l}$ is the Fourier series coefficients of $p_l(t)$, and $L_p \geq BT$ is a positive integer. Let $\tilde{\mathbf{y}}_l^{cs} \in \mathbb{C}^{M_r}$ be the DFT of $\mathbf{y}_l^{cs}$. Define $\mathbf{T}_l$ as a Toeplitz matrix generated by the sequence $\{\rho_{-L_0,l}, \cdots, \rho_{L_0,l}\}$,

$$\mathbf{T}_l = \frac{1}{\sqrt{M_r}} \begin{bmatrix} \rho_{-L_0+N_r-1,l} & \rho_{-L_0+N_r-2,l} & \cdots & \rho_{-L_0,l} \\ \rho_{-L_0+N_r,l} & \rho_{-L_0+N_r-1,l} & \cdots & \rho_{-L_0+1,l} \\ \vdots & \vdots & \ddots & \vdots \\ \rho_{L_0,l} & \rho_{L_0-1,l} & \cdots & \rho_{-L_0+M_r-1,l} \end{bmatrix}, \quad (6)$$

where $L_0 = (N_r + M_r)/2 - 1$. Then under the assumption of ideal bandpass filter, it can be shown [11] that $\tilde{\mathbf{y}}_l^{cs}$ is related the Nyquist sample vector $\mathbf{y}_l$ as

$$\tilde{\mathbf{y}}_l^{cs} = \mathbf{T}_l \mathbf{W} \mathbf{y}_l , \quad (7)$$

where $\tilde{\mathbf{y}}_l^{cs} \in \mathbb{C}^{M_r}$ and $\mathbf{W} \in \mathbb{C}^{N_r \times N_r}$ is a DFT matrix,

$$\mathbf{W}_{k,n} = \sqrt{1/N_r} \exp\left( -j 2\pi (n-1)(k-1-N_r/2)/N_r \right). \quad (8)$$

By stacking the compressive measurement vectors $\tilde{\mathbf{y}}_l^{cs}$ ($l = 1, 2, \cdots, N_a$) into a vector $\tilde{\mathbf{y}}^{cs} \in \mathbb{C}^M$ where $M = M_r N_a$, we have from (4) and (6) that

$$\tilde{\mathbf{y}}^{cs} = \text{diag}(\mathbf{T}_1, \cdots, \mathbf{T}_{N_a})(\mathbf{I}_{N_a} \otimes \mathbf{W})\mathbf{Dz}, \quad (9)$$

where $\text{diag}(\mathbf{T}_1, \cdots, \mathbf{T}_{N_a})$ is a block diagonal matrix, $\mathbf{I}_{N_a}$ is an $N_a \times N_a$ identity matrix and $\otimes$ denotes the Kronecker product. Define $\mathbf{\Phi} \triangleq \text{diag}(\mathbf{T}_1, \cdots, \mathbf{T}_{N_a})(\mathbf{I}_{N_a} \otimes \mathbf{W})\mathbf{D}$. (9) is simplified as

$$\tilde{\mathbf{y}}^{cs} = \mathbf{\Phi z}. \quad (10)$$

In CS theory, the matrix $\mathbf{\Phi}$ is called measurement matrix. It is seen that the QuadCS randomly projects the Nyquist sample space into a low-rate space.

When a fixed QuadCS is used for SAR sampling, $\mathbf{T}_1 = \mathbf{T}_2 = \cdots = \mathbf{T}_{N_a}$. Then the proposed sampling strategy introduces more randomness into the measurement system $\mathbf{\Phi}$ than that by a fixed spreading signal. It is expected that better image reconstruction performance can be obtained.

### 3.2. Fast Imaging

To reconstruct the SAR image, the vector $\mathbf{z}$ is assumed to be sparse in some basis $\mathbf{\psi}$, so that $\mathbf{z} = \mathbf{\psi x}$. We take the wavelet basis for the sparse representation.

Define $\mathbf{A} = \mathbf{\Phi \psi}$ as the sensing matrix. (10) is transformed into

$$\tilde{\mathbf{y}}^{cs} = \mathbf{Ax}. \quad (11)$$

The image reconstruction can be formulated as solving the following $l_1$ minimization problem

$$\hat{\mathbf{x}} = \arg\min_{\mathbf{x}} \left\{ f(\mathbf{x}) = \frac{1}{2} \|\mathbf{Ax} - \tilde{\mathbf{y}}^{cs}\|_2^2 + \lambda \|\mathbf{x}\|_1 \right\}. \quad (12)$$

There are many algorithms to solve (12), in which iterative shrinkage-thresholding algorithm (ISTA) is preferred in this paper. As seen from the algorithm flowchart in [12], at each iteration, the multiplication operation of $\mathbf{A}$ or $\mathbf{A}^H$ with a vector consumes the main computation cost. If the cost is reduced, the fast image reconstruction can be obtained. Note that the matrix $\mathbf{A}$ consists of two parts: $\text{diag}(\mathbf{T}_1, \cdots, \mathbf{T}_{N_a})$ and $(\mathbf{I}_{N_a} \otimes \mathbf{W})\mathbf{D\psi}$. The multiplication of the second part with a vector can be calculated using fast algorithms because of special structures of $\mathbf{W}$, $\mathbf{D}$ and $\mathbf{\psi}$. Then the main cost has to be paid on computing the multiplication of the first part with a vector. It can be shown that the complexity is about $\mathcal{O}(N_0^3 \gamma)$ for $N_r = N_a = N_0$ where $\gamma = M_r / N_r$.

For the proposed sampling strategy, it will be seen that with a proper setting of the spreading signals, $\text{diag}(\mathbf{T}_1, \cdots, \mathbf{T}_{N_a})$ has a structure suitable for fast computation.

Let the Fourier series coefficients of $p_l(t)$ satisfy

$$\rho_{-L_0 + N_r + k, l} = \rho_{-L_0 + k, l}, \quad k = 0, \cdots, M_r - 2. \quad (13)$$

The spreading signal $p_l(t)$ can be designed with an inverse Fourier transform. Define a $N_r \times N_r$ circulant matrix $\mathbf{C}_l$ formulated from sequence $\{\rho_{-L_0, l}, \cdots, \rho_{-L_0 + N_r - 1, l}\}$ as

$$\mathbf{C}_l = \begin{bmatrix} \rho_{-L_0 + N_r - 1, l} & \rho_{-L_0 + N_r - 2, l} & \cdots & \rho_{-L_0, l} \\ \rho_{-L_0, l} & \rho_{-L_0 + N_r - 1, l} & \cdots & \rho_{-L_0 + 1, l} \\ \vdots & \vdots & \ddots & \vdots \\ \rho_{-L_0 + N_r - 2, l} & \cdots & \rho_{-L_0, l} & \rho_{-L_0 + N_r - 1, l} \end{bmatrix}. \quad (14)$$

Let $\mathbf{\Delta} \in \mathbb{R}^{M_r \times N_r}$ be a truncation operation with its elements as

$$\mathbf{\Delta}_{i, j} = \begin{cases} 1, & j = i, i = 1, \cdots, M_r; \\ 0, & \text{else}. \end{cases} \quad (15)$$

Then we have

$$\mathbf{T}_l = \sqrt{1/M_r} \mathbf{\Delta C}_l. \quad (16)$$

Note that the matrix $\mathbf{T}_l$ can be further decomposed as

$$\mathbf{T}_l = \sqrt{1/M_r} \mathbf{\Delta F}_r \text{diag}(\sqrt{N_r} \mathbf{F}_r \mathbf{q}_l) \mathbf{F}_r^H, \quad (17)$$

where $\mathbf{q}_l = [\rho_{-L_0 + N_r - 1, l}, \rho_{-L_0, l}, \cdots, \rho_{-L_0 + N_r - 2, l}]$. We have

$$\begin{aligned} &\text{diag}(\mathbf{T}_1, \cdots, \mathbf{T}_{N_a}) \\ &= \sqrt{N_r / M_r} (\mathbf{I}_{N_a} \otimes \mathbf{\Delta})(\mathbf{I}_{N_a} \otimes \mathbf{F}_r) \text{diag}(\mathbf{Q})(\mathbf{I}_{N_a} \otimes \mathbf{F}_r^H), \end{aligned} \quad (18)$$

where $\text{diag}(\mathbf{Q})$ is a diagonal matrix with its diagonal element from the elements of $\text{vec}(\mathbf{Q})$ and $\mathbf{Q} = [\mathbf{q}_1, \cdots, \mathbf{q}_{N_a}]$. Note that $\mathbf{F}_r$ is a DFT matrix. Then we can compute the multiplication of $\text{diag}(\mathbf{T}_1, \cdots, \mathbf{T}_{N_a})$ with a vector by the fast Fourier transform. It can be shown that by the proposed spreading signals, the computational complexity is about $\mathcal{O}(N_0^2 \log(N_0))$. Similar reasoning is applied to compute the multiplication of $\mathbf{A}^H$ with a vector.

## 4. SIMULATIONS

In our simulations, we use a real SAR image shown in Fig. 2(a) (resolution $0.1 \text{ m} \times 0.1 \text{ m}$) from Sandia National Laboratories as the test image. The SAR image is first used to generate the baseband echoes by the method in [13]. The baseband echoes are upconverted to the IF signals, which are passed through the QuadCS sampling system to obtain the compressive baseband signals. The Nyquist data size is $1024 \times 1024$ and the sub-Nyquist data size in range dimension varies with the compression ratio $\gamma = M_r / N_r$. The Daubechies-4 wavelet basis is taken to sparsely represent the real SAR image. The ISTA with $\lambda = 0.01$ is used to reconstruct the SAR image by the sub-Nyquist data.

The reconstructed images by the proposed scheme with $\gamma = 1/4$ and $\gamma = 1/8$ are given in Figs. 2 (b) and (c). In comparison with original image (Fig. 2 (a)), it is clear that the proposed scheme with only 25% of the Nyquist data achieves the imaging quality of the Nyquist data. For $\gamma = 1/8$ , the reconstructed image (Fig. 2 (c)) also has high quality. However, the enlarged part shows that the sub-Nyquist data is not sufficient. Figs. 2 (d) and (e) shows the reconstructed images by the proposed scheme with fixed spreading signals. The advantages using independent spreading signals are obvious. For comparison, we also simulate the Xampling scheme [7]. From the reconstructed images (Fig. 2 (f) and (g)), we see that the scheme produces poor SAR images for the simulated data.

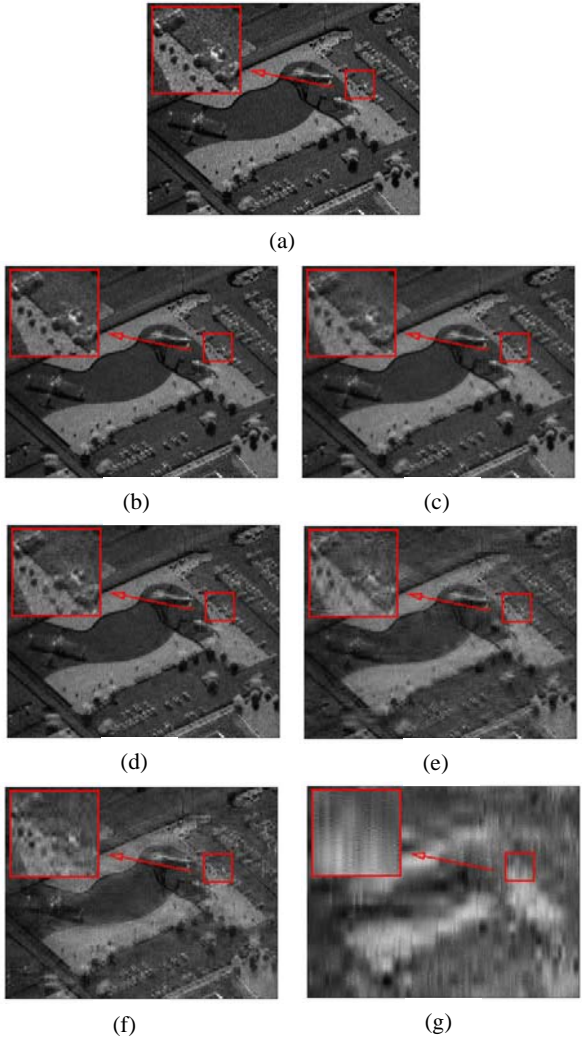

Fig. 2. SAR images (The left-up corner is the enlarged part of the small region). (a) Original image; (b)-(c) QuadCS with independent spreading signals, $\gamma = 1/4$ and $1/8$ ; (d)-(e) QuadCS with fixed spreading signal $\gamma = 1/4$ and $1/8$ ; (f)-(g) Xampling, $\gamma = 1/4$ and $1/8$ .

## 5. CONCLUSION

We presented a QuadCS-based SAR sampling scheme which incorporates independent spreading signals into QuadCS for sampling different SAR echoes. The proposed scheme generates a better sensing matrix and then has better SAR imaging performance. Simulations show that the reconstructed SAR images with only one-fourth of the Nyquist data achieve the image quality similar to that of the classical SAR images with Nyquist samples.

We next discussed fast computation problem in sparse reconstruction of SAR image. We show that with proper setting of the spreading signals, the sensing matrix has structures suitable for fast computation of matrix-vector multiplication, which is common in most reconstruction algorithms. The computational complexity of the ISTA with the sensing matrix is reduced from $\mathcal{O}(N_0^3 \gamma)$ to $\mathcal{O}(N_0^2 \log(N_0))$ .

Future works include theoretical analysis on the sensing matrix and the development of QuadCS SAR with the sub-Nyquist sampling in azimuth dimension.